# A layered single-side readout DOI TOF-PET detector

**L Bläckberg[1], S Sajedi[1], G El Fakhri[1] and H Sabet[1]**
[1]Gordon Center for Medical Imaging, Department of Radiology, Massachusetts General Hospital & Harvard Medical School, Boston, MA, USA

E-mail: LBlackberg@MGH.Harvard.edu, HSabet@MGH.Harvard.edu

**Abstract.** We are exploring a scintillator-based detector with potential of high sensitivity, DOI capability and timing resolution in a single-side readout configuration. Our concept combines: 1) A design with 2+ crystal arrays stacked with relative offset, with inherent DOI information but good timing performance has not been shown with conventional light sharing readout. 2) Single crystal array with one-to-one coupling to the photodetector (PD) array, with good timing performance, but no DOI. We believe the combination, where the first layer of a staggered design is coupled one-to-one to a PD array may provide both DOI and timing. The concept is here evaluated through light transport simulations. Results show that: 1) In terms of DOI one-to-one readout of the first layer allows for accurate DOI extraction using a single threshold relative to the PD sum signal, for up to 4 layers. The number of PD pixels exceeding the threshold corresponds to the layer of interaction. The corresponding approach is not possible for the same geometries with a light sharing readout scheme. 2) When employing a low threshold of 2 optical photons the layered approach with one-to-one readout of the first layer improves timing close to the PD compared to single layer, due to reduced crystal thickness. Single detector timing resolution values of 91, 127, 151 and 164 ps were seen in the 4-layer design with unpolished pixels, compared to 148 ps for single array with one-to-one coupling. 3) For the layered design with light sharing readout, timing improves with increased PD pixel size, with an apparent tradeoff between spatial resolution and timing not observed for the one-to-one coupled counterpart. The combination of straightforward and accurate DOI determination, good timing performance and relatively simple design makes the proposed detector a promising candidate for brain dedicated DOI TOF-PET.

## 1. Introduction

Positron emission tomography (PET) is an imaging modality able to provide functional information related to biological processes in the body which has implications for a number of diseases. For brain imaging in particular, PET can be used to both to detect cancer and aid in diagnosis of neurodegenerative diseases such as dementia [1], Alzheimer's disease [2], and Parkinson's disease [3]. In all these cases early diagnosis is key for good prognosis which emphasizes the value of the availability of high-performance imaging. Furthermore, the significance of the modality is increasing given the continuous development of novel radiopharmaceuticals [4, 5] and analysis methods. In addition, there are currently a large number of ongoing hardware developments driven by the limitations of traditional Whole Body PET (WB-PET) scanners. Examples include high coverage total body PET systems [6] as well as organ specific scanners such as brain dedicated PET [7] and breast dedicated PET [8, 9].

Brain dedicated scanners are motivated by the fact that the resolution and sensitivity of conventional WB-PET is not high enough to accurately image small brain structures [7]. Designing a scanner that targets the brain alone allows to reduce the bore size which will lead to less resolution degradation because of annihilation photon noncollinearity and it will also allow to increase the geometric sensitivity for the brain without excessive use of detector material. There are a number of scanner geometries being explored including small-bore cylindrical systems [10, 11] as well as systems with non-conventional geometries such as the helmet-PET [12, 13], or a wearable PET system [14, 15]. The main purpose of the unconventional geometries is to increase sensitivity compared to cylindrical geometries.

For a scanner configuration with small to medium bore size and/or an unconventional geometry, Depth of Interaction (DOI) capability of the detectors in use is more important compared to in WB-PET due to the increased probability of oblique angles of incidence on the detector face. The severity of the parallax error, manifested as resolution degradation at the edge of the Field of View (FOV), caused by mapping all events to the center of the detector element, increases with crystal thickness, which results in a tradeoff between sensitivity and spatial resolution unless the interaction depth can be determined. High resolution across the whole FOV is especially important in brain imaging due to the off-center location of the cerebral cortex. Currently developed brain dedicated systems employ 2-4 levels of DOI achieved through a number of approaches such as crystal layers with different decay times, varied dopant concentration or varied reflector configurations [7]. In addition to these concepts there are also other approaches for DOI determination including dual-end readout, monolithic crystals with depth dependent light spread, as well as staggered, or relative offset detectors which is the concept that we are exploring further in this work [16, 17] [18] [19].

Another key feature in PET is Time-of-Flight (TOF), which has been shown to improve signal to noise ratio (SNR) in WB-PET by effectively increasing the system sensitivity [20]. The majority of TOF-PET detectors that have been developed are for the application of WB-PET, as this is where the TOF information can have the greatest impact due to the large bore size. For medium bore systems, such as brain dedicated PET, the typical coincidence timing resolution (CTR) of 400 ps FWHM used in WB-PET systems is not good enough to achieve desirable benefit in terms of image quality. However, there has been research showing that TOF can be beneficial also for smaller systems, and it might be favorable to use thinner detectors without DOI but with TOF compared to thicker detectors with DOI but without TOF, as the TOF information can compensate for the loss in geometric sensitivity and DOI degradation is less of an issue for thinner crystals [21]. TOF can also be used to compensate for missing view angles as might be the case in unconventional geometries [22].

We are aiming to develop a detector with both TOF capability and DOI information, as well as high sensitivity and single side readout to reduce system complexity and cost. This is a challenging task given the many trade-offs that typically must be made between these different metrics. Many proposed designs that show good performance have a high level of complexity due to a large number of channels in for example dual-end and side-readout configurations [23-25].

The approach of our novel detector concept is to combine two previously studied designs. The first is the staggered detector consisting of stacked detector arrays relatively offset from each other. This configuration has inherent DOI information due to the differences in light spread between layers. By employing Anger logic to the readout signals pixel maps from individual detector layers will be shifted from one another making 3D positioning possible in a single-side readout configuration. However, this design is typically read out using light sharing schemes and has not been demonstrated to provide good timing resolution. To the best of our knowledge, previous investigation of this detector configuration has not been able to provide good CTR, or timing has simply not be considered as the target application has been small animal PET where the sub-10 cm bore size makes the usefulness of TOF information limited given the CTR achievable using todays detector technologies. The second design is the single detector array one-to-one coupled to a photodetector array, which has been shown to provide superior timing performance compared to its light sharing counterparts, but the design lacks DOI information.

We believe that by combining these concepts into a detector composed of 2 to 4 stacked detector arrays that are each offset from its neighboring layers by half pixel pitch in one dimension, with one-to-one readout of the first detector array can combine desirable characteristics of the two previous designs. Our hypothesis is that such configuration should improve timing performance of the staggered detector compared to the conventional light sharing schemes, as the light will be shared between fewer pixels.

Properties of a scintillator detector system that affects the timing resolution includes intrinsic timing properties of the scintillator material (rise and decay time as well as the light yield), light collection efficiency and light transport from the interaction point to the photodetector, response time, uniformity and noise of the photodetector as well as other downstream electronics. Given the performance of currently available photodetectors, timing is ultimately largely limited by the light generation and transport within the scintillator crystal [26].

In addition, as the DOI characteristics of a scintillator detector is also largely dependent on the spread of scintillation light within the detector, light transport simulations are a relevant first step in assessing the performance characteristics of any scintillator detector configuration.

This paper presents a series of light transport simulations conducted in order to 1) Investigate the potential of the proposed offset detector configuration with one-to-one coupling between the first layer and the photodetector and 2) Compare the performance with staggered detectors with light sharing schemes as well as single layer detectors.

All detectors are compared in terms of DOI performance, timing capability and light collection efficiency.

## 2. Materials and methods

All simulations presented in this work have been conducted using the Monte Carlo code DETECT2000 [27], which is a ray tracing code developed specifically for light transport in scintillator detectors. It should be noted that only optical photons are transported, and hence the gamma interactions themselves were not simulated in this work.

*2.1. Simulated Detector geometries*

All detector configurations in this study are based on Cerium doped Lutetium Yttrium Oxyorthosilicate (LYSO:Ce) with a total crystal thickness of 20 mm. We are investigating using 2 to 4 scintillator array layers, where each layer is offset from its neighboring layers by half pixel pitch in one dimension and the bottom array is coupled one-to-one to a Multi Pixel Photon Counter (MPPC) array with the same pixel pitch, as illustrated in Figure 1. In the 2-layer design each layer is 10 mm thick, in the 3-layer design each layer is 6.67 mm thick and in the 4-layer design each layer is 5 mm thick, totaling 20 mm in all cases. For comparison also the single layer detector with one-to-one coupling has been simulated as well as the staggered detector configuration with readout of the optical photons through light sharing. The readout schemes implemented for crystal/MPPC coupling are illustrated in Figure 2.

Given the target intrinsic transversal resolution of 2 mm our design is based on 2 mm pixel size (2.2 mm pitch) of both the crystal and MPPC arrays. The reflector thickness was chosen as 0.2 mm to match the dead space in the MPPC array. For consistency this reflector thickness was kept for all implemented detector configurations. The light sharing schemes were simulated with three different pixel size combinations: 2.2 mm pixels for both crystal and MPPC pixel pitch with readout scheme 3 (see Figure 2) and a more conventional readout scheme with 4 crystal pixels per MPPC pixel (arrangement 2 in Figure 2). For the latter version both 3.2 and 4.2 mm pixel pitch was used for the MPPC array, corresponding to 1.6 and 2.1 mm pitch for the crystal pixels respectively. For all light sharing schemes, a 1 mm thick light guide was used. Per each detector configuration the quality of the side surfaces of the crystal pixels was varied. Specifics of the implemented detector configurations are summarized in Table 1.

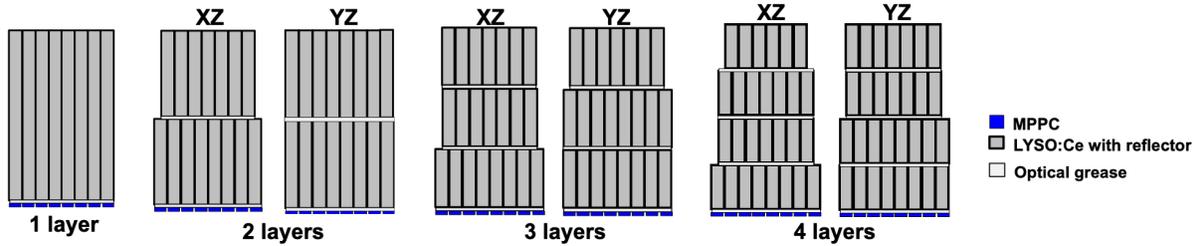

**Figure 1.** Schematic drawing of the detector geometries based on one to one coupling with the photodetector array. In the layered configurations each layer is offset from its neighbor with half pixel pitch in one dimension. The same crystal configurations are explored also with light sharing according to the readout schemes illustrated in Figure 2.

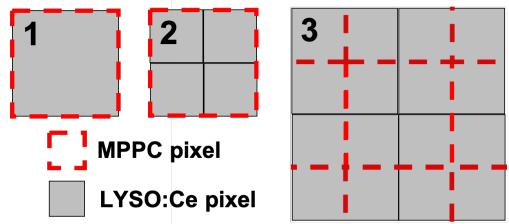

**Figure 2.** Schematic illustration of the arrangements of MPPC pixels with respect to the crystal pixels in the bottom layer of the crystal stack. Three configurations were simulated: 1) MPPC pixels and crystal pixels aligned and of same size. 2) Crystal pixels smaller than MPPC pixels with 4 crystal pixels per MPPC pixel. 3) MPPC pixels and crystal pixels of same size but off set by half pixel pitch in both X and Y directions. Configuration 2 and 3 correspond to light sharing schemes and a 1 mm light guide is used between the crystal and MPPC. Configuration 1 corresponds to one-to-one coupling and no light guide is used.

**Table 1.** Summary of simulated detector configurations.

| Category | Layers | Pixel side surface properties | Pixel pitch (mm) Crystal/MPPC | Crystal-MPPC coupling (See Figure 2) |
|---|---|---|---|---|
| One-to-one coupling | 1,2,3,4 | Polished, Unpolished | 2.2/2.2 | 1 |
| Light sharing | 1,2,3,4 | Polished, Unpolished | 2.2/2.2, 1.6/3.2, 2.1/4.2 | 2,3 |

*2.2. Simulation parameters*

2.2.1 LYSO:Ce crystal arrays

The LYSO:Ce scintillator material was modelled with a refractive index (RI) of 1.82 at the wavelength of maximum emission (420 nm). Consistently with our previous simulation work of this material we were using an optical absorption length of 40 cm, and optical scattering in the crystal bulk was neglected [28, 29]. The entrance and exit surfaces of each pixel were simulated as polished in all cases, and the side surfaces were simulated with two different surface properties: polished and unpolished. The polished surfaces were simulated with the POLISH surface model in DETECT2000, corresponding to 100% specular reflection. The unpolished surface was modeled using the UNIFIED surface model implemented in DETECT2000 [30]. Here 100% specular lobe reflection was used with $\sigma_\alpha$=20°, where the parameter $\sigma_\alpha$ corresponds to the standard deviation of a Gaussian distribution of surface normals around the nominal surface normal. Each crystal pixel was simulated with an external diffuse reflector with a reflection coefficient of 0.98, corresponding to 3 layers of Teflon tape [31]. Transport of optical photons within the reflector material was not considered, and optical photons not subject to reflection were treated as absorbed by the reflector material. The scintillator decay was incorporated in the simulations as a sum of one rising and two decreasing exponentials as outlined in Eq. (1), with $\tau_1$= 21.5 ns, $\tau_2$= 43.8 ns, $\tau_R$= 68 ps, $\alpha_1$= 0.13 and $\alpha_2$= 0.87 [32].

$$f(t) = \left(\frac{1}{\alpha_1\tau_1+\alpha_2\tau_2-\tau_R}\right)\left(\sum_{i=1}^{2} \alpha_i e^{(-\frac{t}{\tau_i})} - e^{(-t/\tau_R)}\right) \qquad (1)$$

2.2.2 Light guide and optical grease

All interfaces between individual crystal layers as well as crystal/light guide and light guide/photodetector surface were coupled with optical grease with a thickness of 100 $\mu$m and a refractive index of 1.47. For the configurations where a light guide was used, this was modeled with a thickness of 1 mm and a refractive index of 1.52.

2.2.3 MPPC array

MPPC arrays with 3 different active pixel areas, 2×2, 3×3 and 4×4 mm$^2$, were implemented. In all cases the dead space between pixels was 0.2 mm. The arrays were simulated with a 100 $\mu$m thick entrance window with a refractive index of 1.57. Optical photons impinging on the active MPPC area were treated as counted with a 45% quantum efficiency. Photons impinging on the dead space were subject to specular reflection with a 30% probability and absorption with 70% probability. This setting was based on our conversations with the manufacturer.

*2.3. Simulation procedure*

Gamma interaction events were simulated throughout the detector thickness in representative pixels, as illustrated in Figure 3. In each location 1000 events were randomly distributed across the pixel cross section within a 0.25 mm slab in the Z-direction. The procedure was repeated every 1 mm in the Z-direction. One gamma event was simulated as a point source of 15000 optical photons emitted isotropically, corresponding to the absorption of 511 keV gamma in LYSO:Ce given its light yield of 27-32k optical photons per MeV.

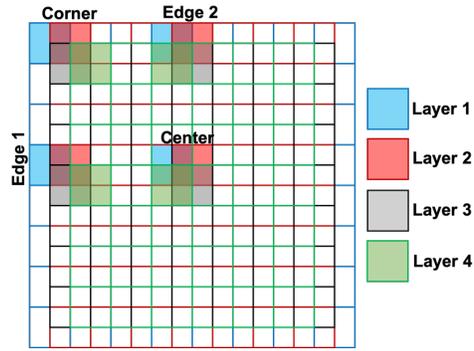

**Figure 3.** Naming convention and pixel locations where gamma events were simulated. Layer 1 is closest to the photodetector plane.

*2.4. Extracted performance metrics*
Per each detector configuration a range of performance metrics were extracted as described in the following subsections.

2.4.1 Depth of Interaction
Depth of interaction (DOI) capability was assessed as a function of interaction position by studying the light signatures of individual gamma events on the photodetector array. Our hypothesis is that DOI determination can be done based on the clear difference between light distributions across the layers, in combination with that the majority of the signals will be measured by the MPPC pixels in direct line of sight of the gamma interaction point. The expected signal patterns from the 4-layer design with one-to-one readout of the first layer are illustrated in Figure 4, and this schematic translates also to the 2- and 3-layer designs if layers are removed from the top.

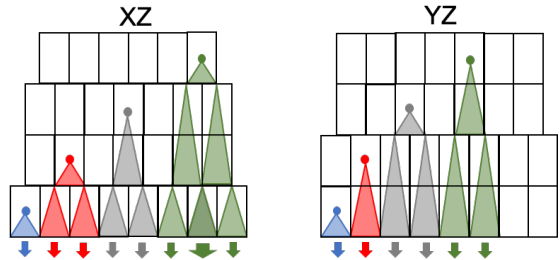

**Figure 4** Illustration of the light spread from each of the layers in the two dimensions. It can be inferred that for layer 1 1x1 MPPC pixels will detect the majority of the light, for layer 2 2x1 pixels will share the majority of the light, for layer 3 2x2=4 pixels and for layer 4 3x2=6 pixels will share the light, where 2 of those will have a higher signal compared to the others.

2.4.2 Light collection efficiency
The light collection efficiency, defined as the fraction of emitted optical photons per event that reached the active area of the photodetector was calculated as a function of event position for each detector configuration.

2.4.2 Timing
The timing resolution in a scintillator detector is governed by the rise and decay times of the scintillator material in use, the light yield and the light collection efficiency, response time of the photodetector as well as time jitter in the downstream electronics. This paper aims to investigate the effects that detector geometry has on the final timing performance, and hence the focus of the analysis is on the timing components related to the scintillator as well as detector geometry. For each optical photon a delay between gamma excitation and optical photon emission was sampled from the probability distribution function given by Eq. 1. In addition, the transit time between emission and detection at the MPPC array was recorded. The two delay components were summed per optical photon to yield a total time delay between gamma excitation and detection at the photodetector.

The influence on timing performance from detection threshold was studied by looking at the time of arrival of the $2^{nd}$, $3^{rd}$, $5^{th}$ and $10^{th}$ optical photon detected by the first MPPC pixel to exceed the threshold defined by the given number of optical photons. For each case a Gaussian function was fitted to the distribution of arrival times, and the centroid (i. e. average delay) and the FWHM (i. e. time jitter in the delay) was extracted and compared across geometries.

## 3. Results

In the following sections the simulated detector configurations are compared across the performance metrics defined in Section 2.4. Throughout this section the default readout scheme is one-to-one coupling between the first crystal layer and the photodetector. LS in the figure legends corresponds to light sharing between the first crystal layer and the photodetector.

*3.1. DOI*

The light signatures predicted in Figure 4 are confirmed by the simulations as manifested in the light distributions shown in Figure 5, Figure 6 and Figure 7. Note that these figures, as well as the following discussion corresponds to the staggered design with one-to-one coupling being the main design in this paper. It can be seen that from events in layer 1, 1 MPPC pixel measures the majority of the light, from layer 2 the major part of the light is shared equally between 2 MPPC pixels, from layer 3 light is divided between 4 MPPC pixels and from layer 4 between 6 MPPC pixels. In the latter case 2 MPPC pixels gets higher signal compared to the other 4. Given that these 1, 2, 4 or 6 dominating signals are always easily separated from the rest of the array the transversal pixel position can be determined from the center of gravity based on these signals, and a transversal resolution corresponding to the pixel size can be achieved. The signal pattern is maintained both in the center, edge and corner for all studied designs, although with slight variations in the numbers.

The fact that the signal strength of the 1, 2, 4 or 6 maximum pixels is always significantly larger than the signals from the rest of the array also implies that a straightforward DOI classification can be performed by simply counting the number of MPPC pixels exceeding a given threshold (relative to the sum signal). The light distributions shown further indicate that a single relative threshold might be used across the whole detector. This approach was tested for each one-to-one coupled configuration using a threshold set at the midpoint between the lower 1 σ bound of all the primary signals across the detector, and the upper 1 σ bound of the remaining signals. The results of this DOI classification are summarized in Table 2. It can be seen that given the smaller error bars (i. e. spread in light distribution across events in a given crystal) for unpolished pixels compared to polished pixels the positioning accuracy is higher in the former case. For unpolished pixels 100% of the events are correctly positioned for the 2- and 3-layer designs, as well as for the major part of the 4-layer design. The latter has reduced positioning accuracy in the $4^{th}$ layer for the center and edge 1 pixels (97.6% and 99.1%), and in the first layer for edge 2 and corner pixels (91.2% and 93.0%).

For polished pixels the accuracy is >99% in all pixels for the 2-layer designs, >96% for the 3-layer designs, and >95% in the two central layers for the 4-layer design. As for unpolished pixels the latter configuration has degraded positioning accuracy in the first and fourth layer. It should be noted however, that if the threshold for central pixels is set without taking into account the detector edge, it can be lowered to 7.8% and achieve positioning accuracy > 99% for all layers. For all designs the mispositioned events were found mainly close to the layer interfaces.

Figure 8 shows a selection of the corresponding light signatures for the staggered design with light sharing readout schemes. Only data for unpolished pixels are shown, but similarly as for the one-to-one readout polished pixels yielded similar average light signatures, but with larger spread between events within a given pixel/layer resulting in larger error bars. Results are also shown only for the 2.2 and the 4.2 mm MPPC pitch, as the 3.2 mm pitch yielded very similar results as the 4.2 pitch given that the same number of crystals per MPPC was used in both cases. From the figure it can be inferred that the outlined single relative threshold approach is not possible for the staggered designs with light sharing, except for two layers with the 2.2 mm MPPC/ 2.2 pixel pitch design. For 2 and 3 layers with 4 crystal pixels per MPPC pixel a single relative threshold could work in the center of the detector, at least for unpolished pixels with smaller error bars, but not toward the edge as signals from layer 1 and 2 will be very similar close to edge 1, and signals from layer 2 and 3 will be very similar close to edge 2. For the 4-layer design with light sharing there are problems already for central pixels for the designs with 4 crystal pixels per photodetector pixel.

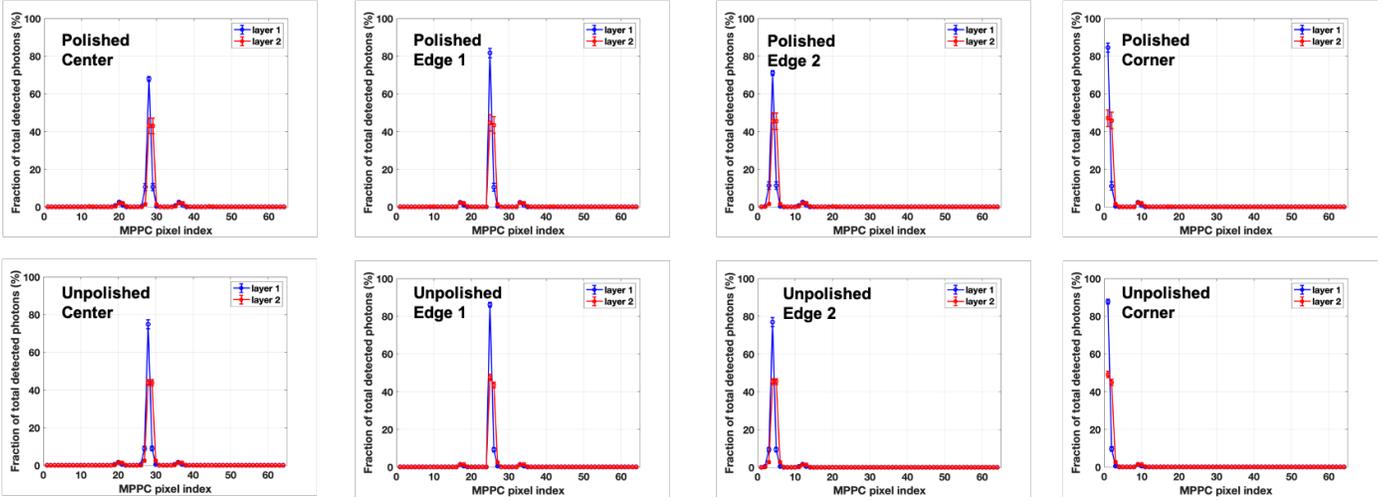

**Figure 5.** Average light distribution per detector layer for the 2-layer design calculated from all events within a given pixel in a given layer. Shown are distributions for 4 representative locations, as described in Figure 3. Values are shown as a percentage of the sum signal, and the error bars correspond to one standard deviation around the mean values. Top: polished crystal pixels, Bottom: Unpolished crystal pixels.

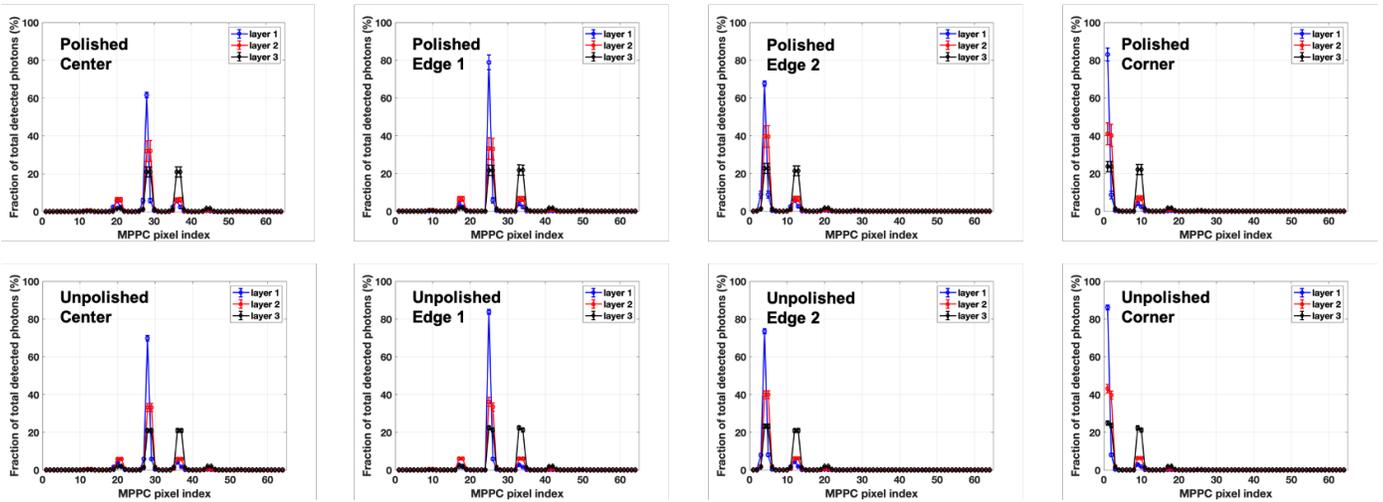

**Figure 6.** Average light distribution per detector layer for the 3-layer design calculated from all events within a given pixel in a given layer. Shown are distributions for 4 representative locations, as described in Figure 3. Values are shown as a percentage of the sum signal, and the error bars correspond to one standard deviation around the mean values. Top: polished crystal pixels, Bottom: Unpolished crystal pixels.

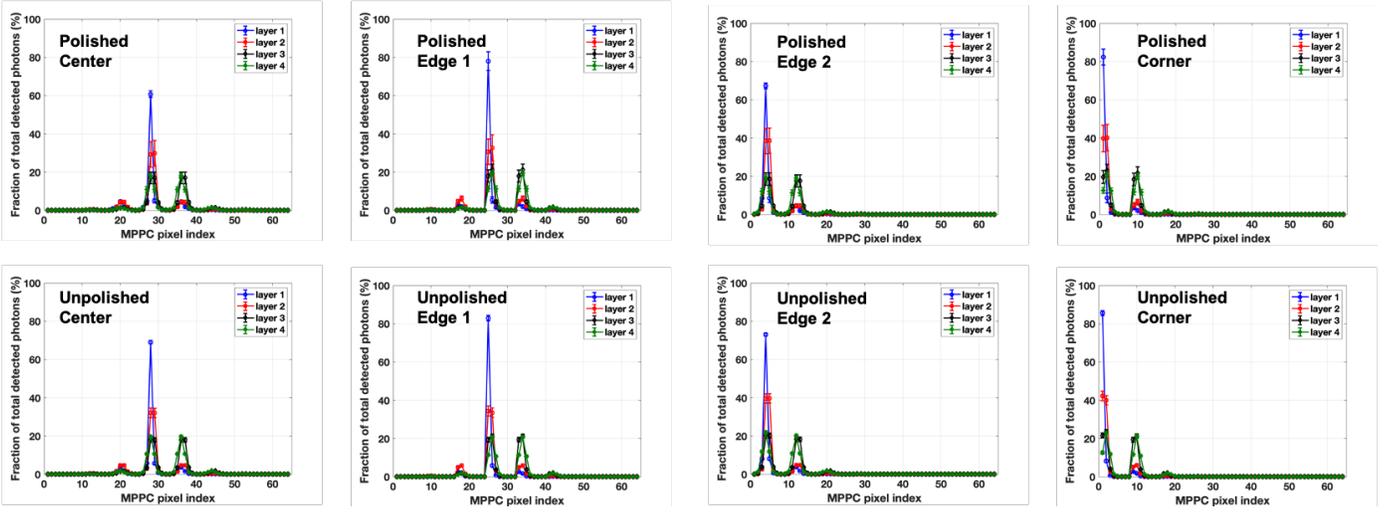

**Figure 7.** Average light distribution per detector layer for the 4-layer design calculated from all events within a given pixel in a given layer. Shown are distributions for 4 representative locations, as described in Figure 3. Values are shown as a percentage of the sum signal, and the error bars correspond to one standard deviation around the mean values. Top: polished crystal pixels, Bottom: Unpolished crystal pixels.

**Table 2.** Positioning accuracy for the staggered design with one-to-one coupled photodetector array. The number of MPPC signals above the threshold determined the layer according to: layer 1 = 1 signal, layer 2 = 2 signals, layer 3 = 3-4 signals and layer 4 = 5-6 signals. The range for the third and fourth layer was chosen over 4 and 6 respectively as positioning accuracy was improved. The threshold was determined as the midpoint between the lower 1 σ bound of the smallest of the 2,4 or 6 signals above the threshold in the top layer and the upper 1 σ bound of the rest of the signals.

| | **Polished** | | **Unpolished** | |
|---|---|---|---|---|
| **Geometry** | Threshold (%) | Positioning Accuracy (%) | Threshold (%) | Positioning Accuracy (%) |
| 2 layers | 26.1 | Center: 100 / 99.9<br>Edge 1: 100 / 99.9<br>Edge 2: 100 / 100<br>Corner: 100 / 100 | 26.4 | Center: 100 / 100<br>Edge 1: 100 / 100<br>Edge 2: 100 /100<br>Corner: 100 / 100 |
| 3 layers | 14.5 | Center: 100 / 98.7 / 96.2<br>Edge 1: 100 / 99.4 / 98.0<br>Edge 2: 98.1 / 100 / 99.0<br>Corner: 97.7 / 100 / 99.35 | 14.4 | Center: 100 / 100 / 100<br>Edge 1: 100 / 100 / 100<br>Edge 2: 100 / 100 / 100<br>Corner: 100 / 100 / 100 |
| 4 layers | 10.1 | Center:100 / 100 / 95.03 / 52.8<br>Edge1: 99.6 / 99.0 / 100 / 68.7<br>Edge2: 74.4 / 100 / 97.9 / 81.4<br>Corner: 83.1 / 98.3 / 100 / 81.4 | 9.5 | Center: 100 / 100 / 100 / 97.6<br>Edge1: 100 / 100 / 100 / 99.1<br>Edge 2: 91.2 / 100 / 100 / 99.9<br>Corner: 93.0 / 100 / 100 / 100 |

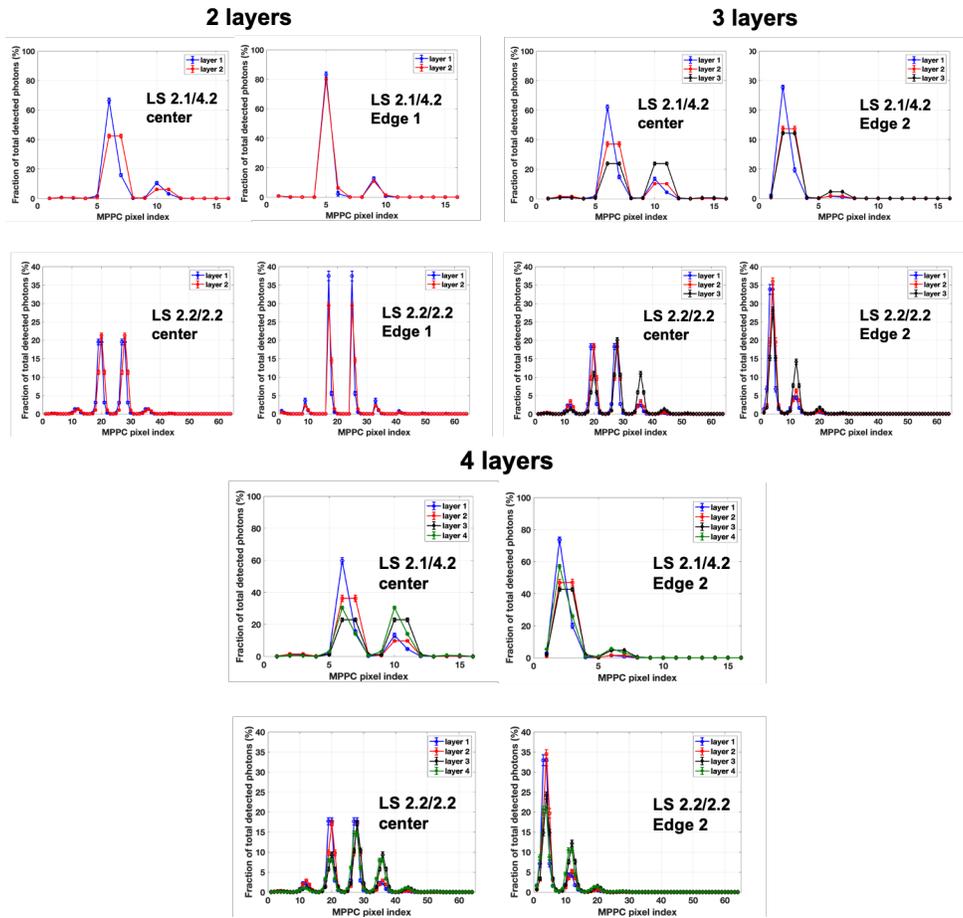

**Figure 8.** Light distributions for the light sharing schemes. Shown are results for a central and an edge pixel for the 2.1/4.2 pitch and the 2.2/2.2 pitch light sharing schemes. The light pattern for the 1.6/3.2 pitch configuration presents the same trends as the 2.1/4.2 design.

*3.2. Light collection efficiency*

Table 3 shows the light collection efficiency (LCE) as an average across the total detector depth for central crystal pixels. Both the single layer and staggered design with one-to-one coupling is compared with light sharing schemes and general trends across all designs are decreased light collection efficiency with number of detector layers and increased light collection efficiency for unpolished compared to polished crystal pixels. The LCE was also found to decrease with distance from the MPPC array in all cases. For polished pixels the LCE is more uniform across the crystal depth compared to unpolished pixels. Our proposed design has higher light collection efficiency compared to the light sharing readout schemes with 2.2 and 3.2 mm MPPC pitch but marginally lower compared to 4.2 mm MPPC pitch. In general, the readout scheme does however not dramatically affect light collection efficiency.

Table 4 compares the total LCE averaged over interaction depth between center, edge and corner for the proposed staggered one-to-one designs. Regardless of number of layers the LCE is higher or equal at the edge and corner compared to central pixels, which is manifested also in the timing performance shown in Figure 13.

**Table 3.** Average light collection efficiency in the center of the array for all detector configurations. Values are shown as percentage of emitted optical photons entering the active areas of the MPPC array. Uncertainties correspond to one standard deviation in the variation between individual events.

|  | One-to-one | | Light sharing 1.6/3.2 | | Light sharing 2.1/4.2 | | Light sharing 2.2/2.2 | |
|---|---|---|---|---|---|---|---|---|
|  | Polished | Unpolished | Polished | Unpolished | Polished | Unpolished | Polished | Unpolished |
| 1 layer | 34.0 ± 0.7 | 69.8 ± 2.4 | 34.0 ± 0.7 | 65.3 ± 3.8 | 34.4 ± 0.7 | 70.0 ± 2.7 | 32.0 ± 0.7 | 64.9 ± 2.4 |
| 2 layers | 32.9 ± 0.7 | 66.4 ± 3.8 | 32.7 ± 0.9 | 60.7 ± 5.6 | 33.6 ± 0.7 | 66.4 ± 4.2 | 31.1 ± 0.7 | 61.8 ± 3.8 |
| 3 layers | 31.8 ± 0.9 | 63.1 ± 5.1 | 31.3 ± 1.1 | 56.4 ± 7.1 | 32.2 ± 0.9 | 62.2 ± 5.6 | 30.0 ± 0.9 | 58.9 ± 4.9 |
| 4 layers | 30.8 ± 1.2 | 60.1 ± 6.2 | 30.1 ± 1.4 | 53.1 ± 8.6 | 31.3 ± 1.2 | 59.9 ± 6.8 | 29.0 ± 1.1 | 56.1 ± 5.9 |

**Table 4.** Variation of LCE in % across the detector cross section for the staggered designs with one to one coupling, and polished and unpolished pixels, respectively.

|  | Polished | | | | Unpolished | | | |
|---|---|---|---|---|---|---|---|---|
|  | Center | Edge 1 | Edge 2 | Corner | Center | Edge 1 | Edge 2 | Corner |
| 2 layers | 32.9 ± 0.7 | 33.3 ± 1.3 | 32.9 ± 0.9 | 33.3 ± 1.3 | 66.4 ± 3.8 | 69.3 ± 5.8 | 66.7 ± 4.0 | 69.8 ± 6.0 |
| 3 layers | 31.8 ± 0.9 | 31.8 ± 1.8 | 32.0 ± 1.3 | 32.2 ± 2.0 | 63.1 ± 5.1 | 65.3 ± 7.8 | 65.3 ± 5.8 | 67.6 ± 8.0 |
| 4 layers | 30.8 ± 1.2 | 31.2 ± 2.1 | 30.9 ± 1.9 | 31.2 ± 2.6 | 60.1 ± 6.2 | 63.5 ± 8.7 | 62.3 ± 7.9 | 65.5 ± 9.7 |

*3.3. Timing*

For each simulated event, the delay from gamma interaction to detection was recorded for the first MPPC pixel to exceed a given threshold of *n* optical photons. The value of *n* was varied from 2 to 10 in order to study the effect of threshold on timing performance.

In Figure 9 the average delay, determined as described in 2.4, as a function of interaction depth is shown for 1, 2, 3 and 4 layers respectively. The one-to-one configuration is here compared with the 3.2 mm MPPC light sharing scheme with 4 crystal pixels per MPPC pixel.

It can be observed that for both readout configurations the delay increases with distance from the MPPC array, and that the layered configurations results in an additional jump in the delay for the second, third, and fourth layer compared to single layer designs. It can further be observed that delay is larger with light sharing compared to one-to-one coupling, and that unpolished crystal pixels shortens the time delay. It is apparent that a layered design will significantly degrade timing performance if the DOI information is not taken into account as the additional jumps in average delay will effectively increase the total FWHM timing resolution. This can be seen as a 250 ps delay difference between events close to the MPPC and events far from the MPPC for the 4-layer design, which can be compared to 150 ps for the single layer with one-to-one coupling (numbers are for *n*=2 and unpolished pixels). However, the fact that DOI can be determined accurately allows to treat each layer independently in terms of timing, leading to significantly shortened delay difference between events close to the MPPC and far from the MPPC in a given layer. Hence, there is potential for improvement compared to single layer as the difference in time of arrival between the top and bottom of one layer can decrease with layer thickness. When looking at Figure 9 it can be seen that this effect should be larger for smaller thresholds.

Figure 10, Figure 11 and Figure 12 shows the FWHM time jitter around the average delay for the 2-, 3- and 4-layer designs respectively, as a function of layer, threshold, photodetector readout scheme and surface quality of crystal pixels. Regardless of design unpolished pixels are yielding better performance compared to polished pixels, and the differences between the readout schemes are larger for higher threshold. For light sharing schemes with 4 crystal pixels per MPPC pixel the timing performance improves with increased photodetector size, as this will yield higher signals from individual MPPC pixels, and hence higher probability to reach the threshold earlier. The light sharing configuration with 2.2 mm pitch and offset photodetector array consistently performs worse compared to the other designs. In general the one-to-one coupling design performs better compared to the light sharing schemes, especially in the layer closest to the MPPC array. However for the 4-layer design (Figure 12) there is timing degradation in the outer layers which is especially pronounced for high thresholds and unpolished pixels.

Figure 13 shows the variation in timing performance in the center, edge and corner of the detector module for the one-to-one staggered designs. It can be seen that in general timing improves close to the edge, especially close to the MPPC array. The differences across the arrays are also found smaller for unpolished pixels and higher thresholds.

Finally, Figure 14 compares the timing performance across all one-to-one coupled designs. It can be seen that unpolished pixels are always better performing compared to polished pixels for the layered designs. For single layer the opposite trend is seen, which is consistent with previous work [26, 33].

Furthermore, it can be seen that increasing the number of layers improves performance in the layer closest to the MPPC array. For a low threshold of $n$=2 the performance of the 3-layer design can be better than the single layer design throughout the detector thickness, which is very encouraging. For 4 layers with the same threshold the performance is only degraded in the outer layer.

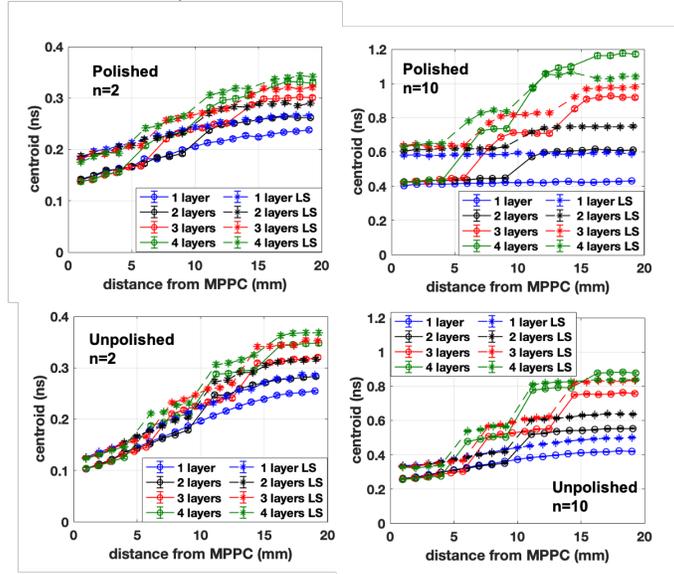

**Figure 9.** Average delay in detection of the n'th optical photon in the MPPC pixel that is first to exceed the threshold defined by n. Here our proposed staggered design is compared to conventional light sharing with 1.6 mm crystal pixel pitch and 3.2 mm MPPC being the intermediate version of the light sharing schemes simulated in this paper. The delay is shown as a function of depth and was determined by Gaussian fit do the distribution of events at the given depth. In this figure only events from a central crystal is shown. Error bars correspond to the 95% confidence interval in the fit parameter. Two different thresholds are shown corresponding to 2 and 10 optical photons, and results are shown for polished and unpolished crystal pixels.

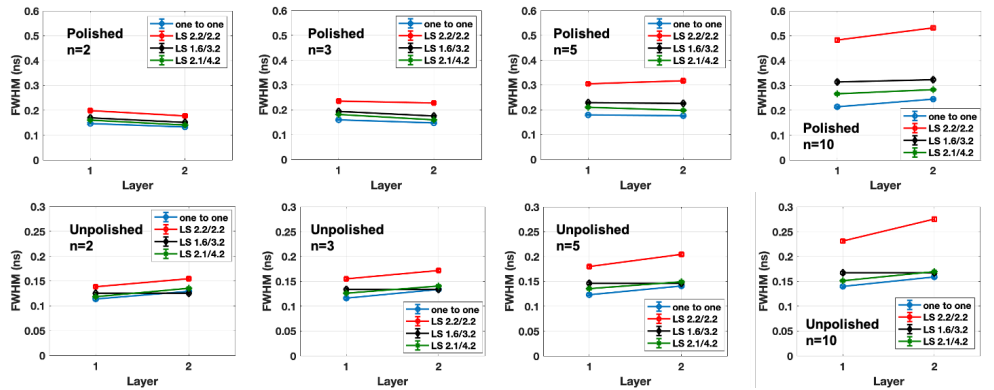

**Figure 10.** Comparison of time jitter across events in a given layer. Shown is data from events within a central pixel in a 2-layer design with different readout schemes. The FWHM values was determined by Gaussian fit to the distribution of events at the given layer. Error bars correspond to the 95% confidence interval in the fit parameter. Four different thresholds are shown corresponding to 2, 3, 5 and 10 optical photons, and results are shown for polished and unpolished crystal pixels.

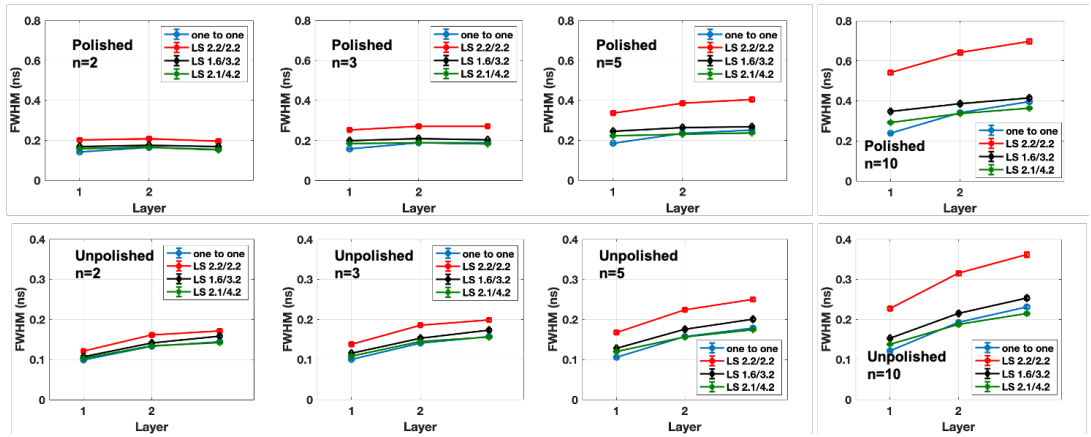

**Figure 11.** Comparison of time jitter across events in a given layer. Shown is data from events within a central pixel in a 3-layer design with different readout schemes. The FWHM values was determined by Gaussian fit to the distribution of events at the given layer. Error bars correspond to the 95% confidence interval in the fit parameter. Four different thresholds are shown corresponding to 2, 3, 5 and 10 optical photons, and results are shown for polished and unpolished crystal pixels.

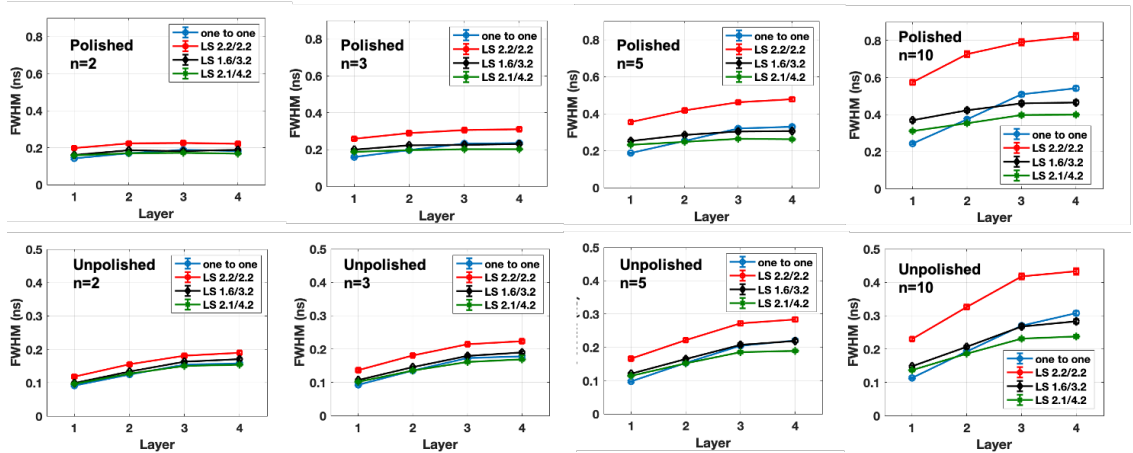

**Figure 12.** Comparison of time jitter across events in a given layer. Shown is data from events within a central pixel in a 4-layer design with different readout schemes. The FWHM values was determined by Gaussian fit to the distribution of events at the given layer. Error bars correspond to the 95% confidence interval in the fit parameter. Four different thresholds are shown corresponding to 2, 3, 5, and 10 optical photons, and results are shown for polished and unpolished crystal pixels.

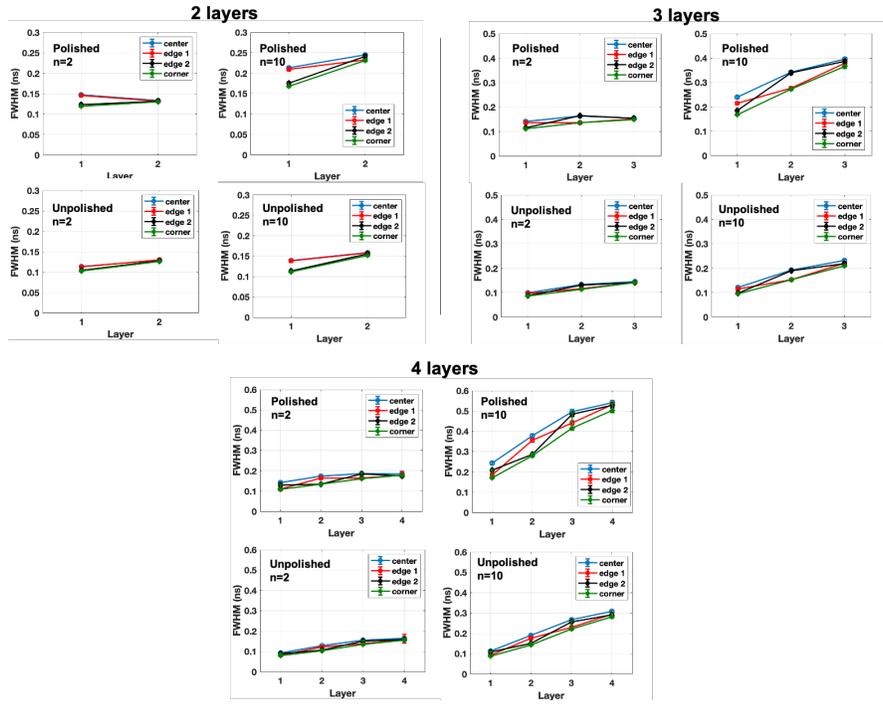

**Figure 13.** Comparison of time jitter across events in a given layer for the proposed 2-, 3- and 4-layer staggered designs with one to one coupling to the MPPC array. Timing performance is compared between central, edge and corner pixels, as defined in Figure 3. The FWHM values was determined by Gaussian fit to the distribution of events at the given layer. Error bars correspond to the 95% confidence interval in the fit parameter. Two different thresholds are shown corresponding to 2 and 10 optical photons, and results are shown for polished and unpolished crystal pixels.

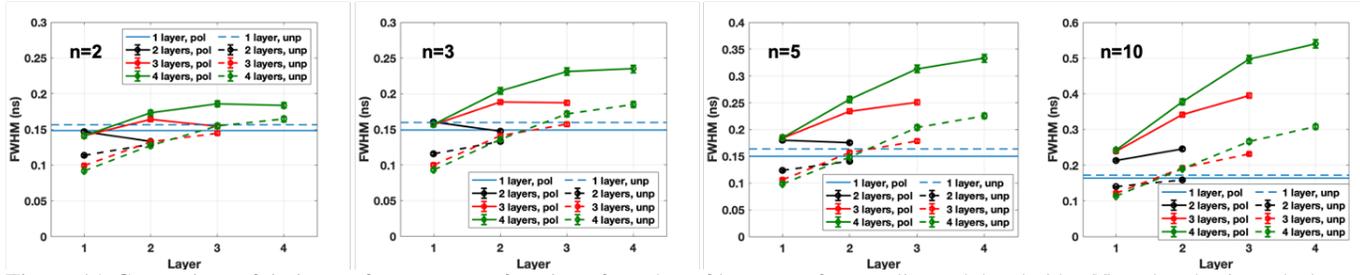

**Figure 14.** Comparison of timing performance as a function of number of layers, surface quality and threshold n. Note that the 4 graphs have different y-axes in order to better highlight the data presented.

## 4. Discussion and conclusions

This paper aims to investigate the performance characteristics of a layered detector design composed of 2+ LYSO:Ce scintillator arrays, each offset from its neighboring layers by half pixel pitch. Similar designs have been previously explored in light sharing configurations. However, in previous work the timing capability of the design has either not been investigated [16, 18, 19] or the presented numbers have been high [17]. As it is well known that for standard single layer detectors timing resolution suffers in the light sharing configuration compared to one-to-one coupling [34], in this work we are exploring the staggered detector design with one-to-one coupling between the first scintillator layer and the photodetector array, and we are also for the first time conducting a thorough evaluation of the timing potential of staggered detector configuration with a light sharing readout scheme. In this regard we have performed an extensive light transport simulation study that compares the proposed one-to-one coupled design with conventional light sharing staggered geometries as well as single layer designs, with and without light sharing. The range of detector configurations have been compared in terms of DOI, LCE and timing performance.

One major advantage of the one-to-one coupled design compared to light sharing is the possibility of straightforward

DOI determination using a single threshold (relative to the sum signal) on the MPPC signals. The 2-, 3- and 4-layered designs explored here where the first layer is coupled one-to-one with the photodetector array, are all characterized by a well-defined layer dependent signal pattern, as illustrated in Figure 5 to Figure 7. For events in the layer closest to the MPPC array the corresponding MPPC pixel measures the majority of the light (60-90% depending on the geometry and pixel surface properties). For events in the second layer the major part of the light is equally shared between the two MPPC pixels in direct line of sight (30-40% per pixel). Correspondingly for events in the third layer light is shared between 4 pixels, with around 20% per pixel. Finally, for events in the fourth layer light is shared between 6 MPPC pixels where two of them measure around 20% each and 4 10% each. The approach of using a single relative threshold to separate these signals from those from the rest of the array was tested (See Table 2) and resulted in 100% positioning accuracy throughout the detector for 2 and 3 layer designs and >91% accuracy for the 4 layer design with unpolished pixels. The degradation for the 4-layer design was mainly towards one of the edges and corner. In the rest of the detector >97% accuracy was achieved. For polished pixels the performance was degraded due to the larger spread between events in a given layer and pixel, as manifested in the error bars in Figure 5 to Figure 7. Here the positioning accuracy was found to be >99% for the 2-layer design and >96% for the 3-layer design. For the 4 layer design the accuracy was >95% in layers 2 and 3 but 52-81% (depending on transversal pixel location) for the fourth layer and 74-100% in the first layer. It should be noted however that these numbers were obtained with the same threshold across the whole array, and if using a position dependent threshold, the accuracy in the center of the detector could be increased to above 99% in all layers.

The corresponding approach was not possible for the light sharing schemes with 4 crystal pixels per MPPC pixel. While a single relative threshold can work for the central part of these detectors for up to 3 layers, the signals from patterns close to the edges tend to overlap making the approach unfeasible. For the 4 layer design single relative threshold is unfeasible already in the center of the detector array. The light sharing design with the same pixel pitch in the crystal and the MPPC works better compared to the other light sharing designs, and the single relative threshold approach may be used throughout the detector for the 2-layer configuration. However, as the signal is spread over more MPPC pixels compared to the one-to-one coupling design they are lower which will make DOI determination more error prone. Problems with accurate positioning at the detector edge has also previously been reported for the 4-layer staggered geometry [19].

The readout scheme used was not found to have a high impact on the light collection efficiency (LCE), see Table 3. For one-to-one readout the average LCE across the whole detector depth was decreasing with 13.9% for unpolished pixels and 9.4% for polished pixels, when going from single layer to 4 layers. Furthermore, the LCE was always found around a factor of 2 higher for unpolished pixels compared to polished pixels with numbers of 30.8-32.9% and 60.1-66.4% respectively. The reason is largely due to less trapping of light by total internal reflection within the pixel volume when the detector sides are roughened. The light collection efficiency was further found to decrease with distance from the MPPC array but was uniform within each layer. The LCE was also found more uniform throughout the detector thickness for polished compared to unpolished pixels. The LCE increased towards the edges and corner of the detector with up to 5, 7 and 9% for 2,3 and 4 layers with unpolished pixels, and around 1.2% for polished pixels. Hence the uniformity in LCE was found better also in the transversal direction for polished compared to unpolished pixels (see Table 4).

In terms of timing we chose to compare the different configurations using the time jitter in the time of arrival of the $2^{nd}$, $3^{rd}$, $5^{th}$ and $10^{th}$ optical photon detected by the MPPC pixel first triggering the mentioned threshold level.

The factors that are reflected in the timing metric are the intrinsic timing properties of the scintillator material given by the rise and decay constants of LYSO:Ce, as well as the travel time from the interaction location to detection by the photodetector. Any other factors affecting the timing performance such as time jitter in the photodetector response, and non-uniform time response between photodetector elements were not included.

Figure 9 shows the average delay between gamma interaction and the detection of the $n^{th}$ optical photon ($n$=2,3,5 and 10) for the 1- to 4-layered designs, with one-to-one coupling or light sharing between the first crystal layer and the photodetector. It can be observed that as expected the delay increases with distance from the photodetector, and for the 4-layer designs with one-to-one coupling there is a delay difference between the first and last layer (for $n$=2) of 180 ps and 230 for polished and unpolished pixels, respectively. The corresponding numbers for 3 layers are 150 and 190 ps, and for 2 layers 90 and 130 ps. For all geometries the delay between the layer closest to the MPPC and the layer at the gamma entrance side of the detector is larger for unpolished compared to polished pixels when a low threshold is used. For $n$=10 the corresponding numbers are 740/620 ps for the 4-layer design, 490/480 ps for 3 layers, and 170/230 ps for 2 layers, and there is no clear trend related to surface properties. It can further be seen that the jump in average delay is significantly smaller between the $3^{rd}$ and $4^{th}$ layer compared to the other discontinuities.

Single detector timing resolution (STR) was determined as the FWHM of the distribution of time delays around the average delay, per layer, as the availability of DOI information can be used to compensate for the delay difference between layers [35].

Figure 10, Figure 11 and Figure 12 shows the STR per layer for the 2-, 3- and 4-layer designs respectively for one-to-one readout compared to light sharing schemes. It can be seen that for light sharing schemes increased photodetector (and crystal) pitch improves timing resolution, and that the benefits of one-to-one coupling of the first layer is seen mainly in the layers closest to the MPPC array, and that especially for the 4-layer design 4-to-1 light sharing can perform better compared to one-to-one coupling in the top layers.

Figure 13 shows that with one-to-one coupling of the first layer timing performance is preserved and can even be improved close to the detector edge and corner, regardless of number of layers and surface finish of the scintillator pixels. Differences across the detector cross-section are larger for polished pixels and larger threshold.

Finally, Figure 14 compares STR per layer for the one-to-one coupling configuration as a function of number of layers and pixel surface finish. The layered configurations are also compared to single layer designs where, in accordance with previous work it is seen that polished crystals is favorable compared to unpolished crystals when it comes to timing performance in pixels with high aspect ratio [26, 33]. For the staggered approach however, the opposite trend is observed, and timing is improved with unpolished crystal pixels. A similar trend change has been seen by others where in the single layer design unpolished pixels are favorable over polished ones when the crystals are shorter (5 cm) while the opposite was true for longer (20 cm) ones [33].

One interesting observation is that timing improves in the crystal layer closest to the MPPC as the total number of layers increases, which corresponds to smaller thickness of individual layers (20 vs 5 mm for the 1- and 4-layer designs respectively). For low threshold of $n=2$ timing is better in all layers for the unpolished 2- and 3-layer designs compared to conventional single layer. For 4 layers however there will be a degradation in the outermost 5 mm layer compared to when considering the FWHM across the whole 20 mm single layer design. With increased threshold the benefit of the layered design compared to single layer is lost, except for the layer closest to the photodetector. When using polished pixels timing is always degraded compared to single layer detectors. From the STR values indicated in Figure 14 the corresponding coincidence timing resolution (CTR) can be calculated as: $CTR = \sqrt{2}\, STR$, which results in values of 210 ps, 160/183 ps, 140/187/205 ps and 129/180/219/232 ps for the 1, 2, 3 and 4 layer configurations, respectively. All numbers are given for a low threshold of $n=2$, unpolished pixels and one-to-one readout of the first layer.

The goal of this ongoing work is to develop a detector with straightforward DOI determination, single-side readout, 2 mm intrinsic resolution and timing performance making it suitable for TOF brain PET. The simulations presented here show promise in that this is achievable with unpolished pixels if a low threshold can be used for time pickoff. Timing can be the same or better compared to a single layer detector with one-to-one coupling for up to 3 layers, and for 4 layers the CTR will be slightly degraded in the fourth layer. When comparing the proposed one-to-one readout with conventional light sharing schemes, the straightforward DOI determination in combination with better timing performance close to the photodetector array makes up for the timing degradation in outer layers associated with the one-to-one coupling approach. When comparing this design to other detector approaches able to provide both DOI and timing information, the practicality and relative simplicity of the concept presented here should be emphasized, as this ultimately will drive the cost of any system level implementation.

Future work entails further optimization with regards to thickness of individual detector layers. As there is a trade-off between DOI and TOF optimization is best done with a given application and scanner geometry in mind.

The trends of improved CTR in the layer closest to the MPPC array with thinner crystals implies the possibility of placing the detector with the photodetector on the entrance side instead of the exit side and in that way achieve good timing and DOI resolution in the entrance part of the detector where the majority of the events take place.

We believe that the proposed design can be interesting both for brain PET as well as other medium sized scanners that requires high spatial resolution and sensitivity and that also can benefit from TOF information, such as breast dedicated PET.

**Acknowledgements**
This work was supported in part by NIH grant R21EB023391. The authors would like to thank Abhishikth Devabhaktuni for help with the data analysis.